\documentclass{aa}  
\usepackage{graphicx,color}
\usepackage{txfonts}

\usepackage{natbib,twoopt}
\usepackage[hyphenbreaks]{breakurl}
\usepackage[breaklinks]{hyperref} 
\bibpunct{(}{)}{;}{a}{}{,}             
\definecolor{cobalt}{rgb}{0.06, 0.2, 0.65}
\hypersetup{
  colorlinks,
  citecolor=cobalt,
  linkcolor=cobalt,
  urlcolor=cobalt,
}
\makeatletter
  \newcommandtwoopt{\citeads}[3][][]{\href{http://adsabs.harvard.edu/abs/#3}%
    {\def\hyper@linkstart##1##2{}%
     \let\hyper@linkend\@empty\citealp[#1][#2]{#3}}}
  \newcommandtwoopt{\citepads}[3][][]{\href{http://adsabs.harvard.edu/abs/#3}%
    {\def\hyper@linkstart##1##2{}%
     \let\hyper@linkend\@empty\citep[#1][#2]{#3}}}
  \newcommandtwoopt{\citetads}[3][][]{\href{http://adsabs.harvard.edu/abs/#3}%
    {\def\hyper@linkstart##1##2{}%
     \let\hyper@linkend\@empty\citet[#1][#2]{#3}}}
  \newcommandtwoopt{\citeyearads}[3][][]%
    {\href{http://adsabs.harvard.edu/abs/#3}
    {\def\hyper@linkstart##1##2{}%
     \let\hyper@linkend\@empty\citeyear[#1][#2]{#3}}}
\makeatother

\def\xmm{{\it XMM-Newton}}

\newcommand{\swift}{{\em Swift}}

\newcommand{\nicer}{{\em NICER}}
\newcommand{\nustar}{{\em NuSTAR}}
\newcommand{\maxi}{{\em MAXI}}

\newcommand{\gtc}{{\em GTC}}
\def\srcfirst{Swift\,J0840.7$-$3516}
\def\src{J0840}

\newcommand{\be}{\begin{equation}}
\newcommand{\en}{\end{equation}}

\def\ltsima{$\; \buildrel < \over \sim \;$}
\def\lsim{\lower.5ex\hbox{\ltsima}}
\def\gtsima{$\; \buildrel $\geq$ \over \sim \;$}
\def\gsim{\lower.5ex\hbox{\gtsima}}

\def\arc{\mbox{$^{\prime\prime}$}}

\def\deg{\mbox{$^{\circ}$}}
\def\nh{\hbox{$N_{\rm H}$}}
\def\flux {\mbox{erg~cm$^{-2}$~s$^{-1}$}}
\def\lum {\mbox{erg~s$^{-1}$}}
\def\fluence {\mbox{erg~cm$^{-2}$}}

\begin{document} 

 \title{Multi-band observations of \srcfirst:\\ a new transient ultra-compact X-ray binary candidate}
  \titlerunning{\srcfirst: a new transient UCXB candidate}
 \authorrunning{F. Coti Zelati et al.}

 \author{F.~Coti Zelati\inst{1,2,3}
	     \and 
	     A.~de~Ugarte~Postigo\inst{4,5}
	     \and
	     T.~D.~Russell\inst{6}
	     \and
        	     A.~Borghese\inst{1,2}
	     \and
	     N.~Rea\inst{1,2}
	     \and
	     P.~Esposito\inst{7,8} 
	     \and \\
	     G.~L.~Israel\inst{9}
	     \and
	     S.~Campana\inst{3}
 	     }

 \institute{Institute of Space Sciences (ICE, CSIC), Campus UAB, Carrer de Can Magrans s/n, E-08193, Barcelona, Spain\\
            \email{cotizelati@ice.csic.es}
       	   \and
           Institut d'Estudis Espacials de Catalunya (IEEC), Carrer Gran Capit\`a 2--4, E-08034 Barcelona, Spain
           \and
	   INAF--Osservatorio Astronomico di Brera, Via Bianchi 46, I-23807 Merate (LC), Italy
           \and
           Instituto de Astrof\'isica de Andaluc\'ia (IAA-CSIC), Glorieta de la Astronom\'ia s/n, E-18008 Granada, Spain
           \and
           DARK, Niels Bohr Institute, University of Copenhagen, Lyngbyvej 2, DK-2100 Copenhagen {\O}, Denmark
           \and
           INAF--Istituto di Astrofisica Spaziale e Fisica Cosmica, Via U.\,La Malfa 153, I-90146 Palermo, Italy
           \and
           Scuola Universitaria Superiore IUSS Pavia, Palazzo del Broletto, piazza della Vittoria 15, I-27100 Pavia, Italy
           \and
	   INAF--Istituto di Astrofisica Spaziale e Fisica Cosmica di Milano, via A.\,Corti 12, I-20133 Milano, Italy
	   \and
	   INAF--Osservatorio Astronomico di Roma, via Frascati 33, I-00078 Monteporzio Catone, Italy
           }

\date{Received 16 February 2021 / Accepted 19 April 2021}

\abstract{We report on multi-band observations of the transient source \srcfirst, which was detected in outburst in 2020 February by the \emph{Neil Gehrels Swift Observatory}. The outburst episode lasted just $\sim$5 days, during which the  X-ray luminosity quickly decreased from $L_X\approx3\times10^{37}d^2_{10}$\,\lum\ at peak down to  $L_X\approx5\times10^{33}d^2_{10}$\,\lum\ in quiescence (0.3--10\,keV; $d_{10}$ is the distance to the source in units of 10\,kpc). Such a marked and rapid decrease in the flux was also registered at UV and optical wavelengths. In outburst, the source showed considerable aperiodic variability in the X-rays on timescales as short as a few seconds. The spectrum of the source in the energy range 0.3--20\,keV was well described by a thermal, blackbody-like, component plus a non-thermal, power law-like, component and it softened considerably as the source returned to quiescence. The spectrum of the optical counterpart in quiescence showed broad emission features mainly associated with ionised carbon and oxygen, superposed on a blue continuum. No evidence for bright continuum radio emission was found in quiescence.
We discuss possible scenarios for the nature of this source and show that the observed phenomenology points to a transient ultra-compact X-ray binary system.}
   
\keywords{methods: data analysis --- methods: observational  --- techniques: spectroscopic -- X-rays: binaries ---  X-rays: individual (Swift\,J0840.7$-$3516)}

\maketitle

\section{Introduction}
\label{sec:intro}

Low-mass X-ray binaries (LMXBs) are systems where a compact object (a neutron star, NS, or a stellar-mass black hole, BH) accretes matter from a low-mass ($M\lesssim1M_\odot$) donor star \citep{frank2002}. 
Many LMXBs spend most of their time in quiescence with little or no mass accretion taking place, but they can undergo sporadic accretion outbursts where the X-ray luminosity can increase up to 5--6 orders of magnitude above quiescence, up to $L_{\mathrm{X}} \sim 10^{38}-10^{39}$\,\lum. The quiescent phase can be as long as decades, whereas the outbursts typically last from weeks to months, and even years \citep[e.g.][]{tetarenko2016}. Such a transient behaviour is commonly ascribed to thermal-viscous instabilities in the accretion disk around the compact object (e.g. \citealt{hameury2016}; for a review, see \citealt{hameury20}).

Ultra-compact X-ray binaries (UCXBs) are a sub-class of LMXBs that are characterised by orbital periods shorter than 80\,min and a hydrogen-deficient donor star, such as a non-degenerate helium star or a white dwarf (\citealt{nelson1986,savonije1986}; for a review, see \citealt{nelemans2010}).  Currently, there are 24 known UCXBs with a measured orbital period and a further 15 systems have been classified as candidate UCXBs \citep{sazonov2020,ng21}.
To date, all confirmed UCXBs have been found to harbour a NS accretor, while for some candidates the nature of the accretor has not been established conclusively \citep[e.g.][]{bahramian2017}. 

Due to the peculiar nature of the donor star in UCXBs, the chemical composition of the accreting material in these systems is expected to be significantly different from that of LMXBs harbouring main sequence donor stars. In fact, observations have shown that the disk material in UCXBs contains elements such as helium, carbon, and oxygen (and in some cases neon and magnesium) in large relative amounts (e.g. \citealt{nelemans2004,nelemans2006,werner2006}). On the other hand, hydrogen is typically the most abundant element in the majority of LMXBs.

On 2020 February 5 at 06:35:51 UT, the Burst Alert Telescope (BAT) on board of the \emph{Neil Gehrels Swift Observatory} 
triggered and located a burst of gamma rays \citep{evans20}. Observations with the \swift\ X-ray 
Telescope (XRT) and the Gas Slit Camera (GSC) on board of the \emph{Monitor of All-sky X-ray Image} (\maxi) 
revealed an X-ray counterpart to the burst \citep{niwano20}, showing considerable flaring activity over the 
subsequent hours, as detected using the \swift/XRT and the \emph{Neutron Star Interior Composition Explorer} 
(\nicer) \citep{iwakiri20}. These properties and the source proximity to the Galactic plane (Galactic latitude of $b\simeq4.0$\deg) 
cast doubt on an interpretation in terms of a `canonical' gamma-ray burst of extragalactic origin, suggesting 
a transient source within the Galaxy instead. The source was dubbed \srcfirst. Observations with ground-based telescopes uncovered an optical counterpart with a rapidly fading emission during the first $\sim$\,20\,h after the initial burst detection \citep{melandri20,malesani20}. 
 
This paper presents the results of follow-up X-ray, UV, optical, and radio observations of \srcfirst, aimed at identifying the nature of this transient. We show that the phenomenology of this source is consistent with what is expected from an UCXB. The paper is structured as follows: we describe the observations and the data analysis in Section\,\ref{sec:data} and report the results in Section\,\ref{sec:results}. A discussion and conclusions follow in Sections\,\ref{sec:discussion} and\,\ref{sec:conclusions}.

\begin{table*}
\scriptsize
\caption{
\label{tab:observations}
Log of pointed X-ray observations, spectral parameters, and X-ray fluxes of \src. The spectra were fitted using an absorbed blackbody plus power-law model, fixing the column density to  $\nh=4.6\times10^{21}$\,cm$^{-2}$ and the photon index to $\Gamma=1.77$ (see Section\,\ref{sec:spectra} for more details).}
\resizebox{2.05\columnwidth}{!}{
\centering
\begin{tabular}{ccccccccccc}
\hline\hline
Instrument\tablefootmark{a}	&Obs.ID	&Start	&Stop							&Exposure	&Count rate\tablefootmark{b}		&$kT_{{\rm BB}}$	&$R_{{\rm BB}}$\tablefootmark{c}	&$F_{X, {\rm obs}}$\tablefootmark{d} 		&$F_{X, {\rm unabs}}$\tablefootmark{d} & Th. fraction\tablefootmark{e}\\
			&		& \multicolumn{2}{c}{YYYY-MM-DD hh:mm:ss (TT)} & (ks)			& (counts s$^{-1}$)	& (keV)			& (km)	& \multicolumn{2}{c}{($\times10^{-11}$\,\flux)} & (\%) \\
\hline
\swift/XRT (WT) & 00954304000 & 2020-02-05 06:38:33 & 2020-02-05 06:40:03 & 0.1  & 30.5$\pm$0.6    & 1.55$\pm$0.06  	& $4.7_{-0.5}^{+0.3}$ 	& 184$\pm$7     & 216$\pm$4     & 57$\pm$5		 \\
\swift/XRT (WT) & 00954304001 & 2020-02-05 07:52:25 & 2020-02-05 16:11:53 & 0.8  & 22.8$\pm$0.2    & 1.16$\pm$0.02  	& 6.0$\pm$0.3	& 123$\pm$2     & 152$\pm$2     & 45$\pm$2		 \\
\swift/XRT (PC) & 00954304001 & 2020-02-05 10:00:19 & 2020-02-05 16:17:53 & 2.7  & 0.52$\pm$0.01   & 0.48$\pm$0.03  	& $4.2_{-0.6}^{+0.4}$ 		& 3.0$\pm$0.2   & 4.2$\pm$0.2   	& 25$\pm$5	  \\
\nicer/XTI      & 2201010101  & 2020-02-05 14:43:34 & 2020-02-05 21:09:00 & 1.3  & 2.15$\pm$0.05   	& --             		&					& --            & --        & --   \\
\swift/XRT (PC) & 00954304002 & 2020-02-06 11:13:26 & 2020-02-06 14:40:52 & 1.0  & 0.061$\pm$0.008 & 0.32$\pm$0.04  	& $4.3_{-0.6}^{+0.8}$ & 0.44$\pm$0.07 & 0.7$\pm$0.1   	& 28$\pm$12  \\
\nicer/XTI      & 2201010102  & 2020-02-06 15:27:10 & 2020-02-06 23:20:50 & 1.8  & 3.78$\pm$0.05   	& --             		&	& --            & --        & --    \\
\swift/XRT (PC) & 00954304005 & 2020-02-06 22:29:17 & 2020-02-07 03:02:50 & 3.3  & 0.176$\pm$0.007 & 0.25$\pm$0.05  	& $7.6_{-2.7}^{+1.9}$ 	& 0.90$\pm$0.06 & 1.4$\pm$0.1   & 15$\pm$6 	 \\
\nicer/XTI      & 2201010103  & 2020-02-07 00:44:31 & 2020-02-07 21:03:51 & 7.2  & 1.52$\pm$0.02   & --             			&	& --            & --         & --   \\
\swift/XRT (PC) & 00954304003 & 2020-02-07 14:12:57 & 2020-02-07 16:17:52 & 2.8  & 0.084$\pm$0.006 & 0.30$\pm$0.03  	& $4.4_{-1.0}^{+0.7}$	& 0.34$\pm$0.03 & 0.53$\pm$0.04 	  & 31$\pm$7 \\
\nicer/XTI      & 2201010104  & 2020-02-08 03:02:14 & 2020-02-08 23:52:00 & 2.7  & 1.04$\pm$0.03   & --            			&	& --            & --         &  -- \\
\nustar/FPMA    & 90601304002 & 2020-02-08 07:01:09 & 2020-02-09 06:56:09 & 42.4 & 0.050$\pm$0.001\tablefootmark{f} & 0.34$\pm$0.05  	& $2.7_{-0.7}^{+2.1}$ 	& 0.25$\pm$0.03 & 0.38$\pm$0.04 	& 29$\pm$9  \\
\nustar/FPMB    & 90601304002 & 2020-02-08 07:01:09 & 2020-02-09 06:56:09 & 41.5 & 0.045$\pm$0.001\tablefootmark{f} & 0.34$\pm$0.05  	& $2.7_{-0.7}^{+2.1}$	& 0.25$\pm$0.03 & 0.38$\pm$0.04 	 & 29$\pm$9  \\
\swift/XRT (PC) & 00954304004 & 2020-02-08 10:55:58 & 2020-02-08 17:36:53 & 3.8  & 0.060$\pm$0.004\tablefootmark{f} & 0.34$\pm$0.05  	& $2.7_{-0.7}^{+2.1}$	& 0.25$\pm$0.03 & 0.38$\pm$0.04 	& 29$\pm$9 \\
\nicer/XTI      & 2201010105  & 2020-02-09 01:08:35 & 2020-02-09 23:06:40 & 12.7  & 0.86$\pm$0.01     & --             		& 		 	& --            	& --            				& --	\\
\swift/XRT (PC) & 00954304007 & 2020-02-09 07:33:15 & 2020-02-09 17:18:53 & 4.1  & 0.029$\pm$0.003 & 0.12$\pm$0.07  	& $10.1_{-4.8}^{+3.2}$	& 0.18$\pm$0.02  & 0.34$\pm$0.01 & 14$\pm$9	 \\
\swift/XRT (PC) & 00954304008--067 & 2020-02-10 13:52:00 & 2021-02-10 17:28:16 & 207.2  & 0.0028$\pm$0.0001\tablefootmark{g} & 0.11$\pm$0.01  & $13.5_{-3.6}^{+5.5}$	& 0.013$\pm$0.001  & 0.035$\pm$0.006 		& 46$\pm$6  \\ 
\hline
\end{tabular}}
\tablefoottext{a}{The instrumental setup is indicated in brackets: PC = photon counting, WT = windowed timing.}
\tablefoottext{b}{Average net count rate. It is in the 0.3--10\,keV energy range, except for \nicer\ (0.5--5\,keV) and \nustar\ (3--20\,keV).}
\tablefoottext{c}{The blackbody radius was calculated assuming a distance of 10\,kpc.}
\tablefoottext{d}{All fluxes are in the 0.3--10\,keV energy range.}
\tablefoottext{e}{The thermal fraction was evaluated as the ratio between the unabsorbed flux of the blackbody component and that of the total emission over the energy range 0.3--10\,keV.}
\tablefoottext{f}{Datasets of these observations were fitted together.}
\tablefoottext{g}{Datasets of these observations (59 in total) were merged.}
\end{table*}

\section{Observations and data reduction}
\label{sec:data}

\subsection{X-ray observations}
\label{sec:xray}

Table\,\ref{tab:observations} reports a journal of all pointed X-ray observations of \srcfirst\ (\src) analysed in this work. In the following, we describe the data processing and analysis. 
All photon arrival times were referred to the Solar System barycenter using the position derived from data taken with the X-Shooter spectrograph mounted on the Very Large Telescope (VLT), 
R.A. = 08$^\mathrm{h}$40$^\mathrm{m}$40$\fs$94, decl. = --35$^{\circ}$16$^{\prime}$25$\farcs$1 (J2000.0; uncertainty of 0.2\arc; \citealt{malesani20}). The spectral analysis was performed 
using the {\sc xspec} fitting package \citep{arnaud96}. Hereafter, all uncertainties are quoted at a confidence level (c.l.) of 1$\sigma$, unless otherwise stated.

\subsubsection{Swift/BAT}
The alert that led to the discovery of \src\ started with an on-board BAT image trigger (Id. 954304, \citealt{evans20}) and the prompt spacecraft slew. The spectral products and the other pieces of information discussed in Section\,\ref{BAT:analysis} were all obtained from a standard analysis of the BAT event data using the instrument-specific tasks in the \textsc{FTools} software package v6.28.

\subsubsection{Swift/XRT}

The \swift/XRT \citep{burrows05} monitored extensively \src\ since the burst trigger. Observations were performed with a daily cadence during 
the first two weeks, about once per week from late February until the end of August 2020 and once every two weeks since then and until one year after the discovery of the source. Except for the two 
prompt post-burst observations in the windowed timing mode (WT; readout time of 1.8\,ms), all data were acquired in the photon 
counting mode (PC; 2.51\,s). The data sets were processed using standard screening criteria. We extracted source 
event lists and spectra using a rectangular region of 40$\times$20 pixels centred on the source for WT-mode data and a circle with a radius of 20 pixels (1 XRT pixel $\simeq$ 2.36 arcsec) for PC-mode data. Background event lists and spectra were collected using a 40$\times$20 pixels box far from the source for WT-mode data and an annulus centred on source with radii of 40 and 80 pixels (free of sources) for PC-mode data. Background-subtracted spectra were extracted separately for the observations between 2020 February 5--9, while a combined spectrum was created from all the subsequent pointings, where the source attained a steady X-ray intensity level (total exposure time of $\sim$207.2\,ks). All spectra were then grouped to contain a minimum of 20 counts per energy bin.

\subsubsection{NuSTAR}

\nustar\ \citep{harrison13} observed \src\ on 2020 February 8--9 for an on-source exposure time of $\sim$42\,ks. Data were processed and analysed 
using \textsc{nustardas} v. 2.0.0 and \textsc{caldb} v.\,20200813. For both focal plane modules (FPMs), the source signal-to-noise ratio 
(S/N) was maximum over the 3--20\,keV energy range (S/N$\sim$34 for the FPMA and $\sim$33 for the FPMB). The subsequent 
analysis was thus limited to this energy interval. Source photons were collected within a circle of radius 80 arcsec, 
while background photons were extracted from a circle of the same size far from the source and on the same detector.  
We extracted background-subtracted light curves and spectra using 
\textsc{nuproducts}. Light curves from each FPM were then combined to increase the S/N. Spectra of both FPMs were grouped to have 
at least 50 counts per energy channel.

\subsubsection{NICER}

The X-ray Timing Instrument (XTI) on board \nicer\ \citep{gendreau12} monitored \src\ intensively over the first week since the burst trigger. We processed and screened the datasets using \textsc{nicerdas} v7a and standard criteria. We additionally extracted the light curves of all observations in the range 12--15\,keV with a time bin of 1\,s, and removed instances of background flares by applying 
intensity filters.  
We estimated the background count rate and extracted the background spectra for each observation using the tool {\sc nibackgen3C50} (v6f). Emission from the source was detectable over the background only during the first five observations and over a narrow energy range in all cases (0.5--5\,keV in obs.ID 2201010102 and 0.8--3\,keV in the other four observations), precluding an adequate modelling of the spectra and meaningful constraints on the spectral parameters. Hence, we refrain here from performing spectral analysis of these data.

\subsection{UV and optical observations}
\label{sec:opt}

\subsubsection{\swift/UVOT}

The \swift\ Ultraviolet and Optical Telescope (UVOT; \citealt{roming05}) observed \src\ all along the outburst using different UV and optical filters. Source photons were extracted adopting a circle of radius 3 arcsec centred on the position of the target. Background photons were collected from a closeby circle of radius 10 arcsec. Photometry was carried out for all observations using the {\sc uvotsource} task, applying aperture corrections.

\subsubsection{GTC}

The Gran Telescopio Canarias (GTC) observed \src\ using the 
Optical System for Imaging and low-Intermediate Resolution Integrated Spectroscopy (OSIRIS; \citealt{cepa00}). An imaging visit was performed starting on 2020 February 18 at 23:46:31 UTC (MJD 58897.991), and consisted of $3\times80$\,s images in each of the $g'$, $r'$ and $i'$ filters. The source was observed at high airmass (2.3) and in poor seeing conditions (between 1.4$^{\prime\prime}$ and 2$^{\prime\prime}$). 

Following this observation, we performed spectroscopy starting on 2020 February 20 at 23:55:20 UTC (MJD 58899.997). The observations consisted of $3\times900$\,s exposures using the grism R1000B and a slit width of 1.0$^{\prime\prime}$, covering the wavelength range from 3700 to 7800\,\AA\ with a resolution of 9.18\,\AA. The seeing was about 1.3$^{\prime\prime}$. The spectrophotometric calibration was performed with respect to the G191-B2B reference star. The reduction and calibration of the spectra were performed with custom scripts based on the \textsc{iraf} package. The spectra were then combined and averaged to increase the S/N.

\subsection{Radio observations}
\label{sec:radio}

The Australia Telescope Compact Array (\emph{ATCA}) observed the field of \src\ on 2020 February 11, between 14:50 and 19:10 UT (MJD 58890.618--58890.799), for a total on-source time of $\sim$4\,hr (project code CX458). The radio data were recorded simultaneously at central frequencies of 5.5 and 9\,GHz, with 2\,GHz of bandwidth at each frequency. Each frequency band was comprised of 2048 1-MHz channels. PKS\,1934$-$638 was used for bandpass and flux calibration, while PKS\,B0826$-$373 was used as the nearby phase calibrator. Data were edited for radio frequency interference, weather, and system issues, before being calibrated and imaged following standard procedures in the Common Astronomy Software Application (\textsc{casa}; \citealt{mcmullin07}). The field was imaged using a natural weighting (a Briggs robust parameter of 2) to maximise the sensitivity of the observations. The source was not detected in our radio observations. The reported 3-$\sigma$ upper-limit on the radio flux density was determined as 3 times the local rms over the source position.

\begin{table}
\scriptsize
\caption{
\label{tab:photometry}
Log of UV and optical photometric observations and source magnitudes or upper limits.}
\centering
\begin{tabular}{cccc}
\hline\hline
Instrument(Filter)	&Start	&Exposure & Magnitude\tablefootmark{a}	\\
&			(TT)		(s)		& \\
\hline
\swift/UVOT($UVW2$)    & 2020-02-05 06:38:30   & 246 	& 17.32$\pm$0.11  \\
\swift/UVOT($UVW2$)   & 2020-02-05 07:52:30   & 261   	& 17.02$\pm$0.10  \\
\swift/UVOT($UVW2$)   & 2020-02-05 08:13:46   & 400   	& 17.32$\pm$0.10  \\
\swift/UVOT($UVW2$)   & 2020-02-05 09:59:37   & 197  	& 18.44$\pm$0.19  \\
\swift/UVOT($UVW2$)   & 2020-02-05 11:14:57   & 598  	& 18.51$\pm$0.13  \\
\swift/UVOT($UVW2$)   & 2020-02-05 14:15:21   & 93  	& $>$18.1  \\
\swift/UVOT($UVW2$)   & 2020-02-05 16:11:52   & 358 	& $>$19.2  \\
\swift/UVOT($UVW2$)   & 2020-02-06 11:13:28   & 548 	& $>$19.3  \\
\swift/UVOT($UVW2$)   & 2020-02-06 22:29:06   & 3372 	& 20.27$\pm$0.19  \\
\swift/UVOT($UVW2$)   & 2020-02-07 14:13:01   & 912 	& $>$19.9  \\
\swift/UVOT($UVW2$)   & 2020-02-08 10:56:02   & 1250 	& $>$20.1  \\
\swift/UVOT($UVW2$)   & 2020-02-09 07:33:19   & 4041    & $>$20.5 \\
\hline
\gtc/OSIRIS($r$) & 2020-02-18  23:47:40  & 3$\times$80 & $\sim$21    \\
\hline
\end{tabular}
\newline
\tablefoottext{a}{For \swift/UVOT, magnitudes are reported only for observations performed in outburst. Values are in the Vega system. 5$\sigma$ upper limits are quoted in case of non-detection.}
\end{table}

\section{Results}
\label{sec:results}

\subsection{BAT characterization of the burst}\label{BAT:analysis}

An image trigger is generated when the BAT detects a rate increase and then finds a point source in a readily-performed image reconstruction of the sky. In the case of \src, the point source had a S/N = 10.5 and was found at the position R.A. = 08$^\mathrm{h}$40$^\mathrm{m}$40$\fs$1, decl. = --35$^{\circ}$16$^{\prime}$02$\farcs$9 (90\% error radius: 2.1\,arcmin).
The T$_{90}$ duration (the time interval containing 90\% of the counts), as computed by the Bayesian blocks algorithm {\sc battblocks}, was $210\pm20$\,s. 

Several simple models can describe the BAT spectrum (in the T$_{90}$ duration range) of \src\ in the 15--150\,keV energy range \citep{evans20,stamatikos2020}. For example, a power law (photon index of $\Gamma=2.3\pm0.1$,
reduced chi squared $\chi_r^2=1.26$ for 56 degrees of freedom, d.o.f.), an optically-thin thermal bremsstrahlung ($kT=42\pm6$\,keV, $\chi_r^2=1.06$ for 56 d.o.f.), a blackbody (temperature of $kT_{\rm BB}=8.7\pm0.5$\,keV, $\chi_r^2=1.12$ for 56 d.o.f.), or a cut-off power-law ($\Gamma=0.3\pm0.7$, cut-off energy of $E_{\mathrm{c}}=19^{+10}_{-6}$\,keV, $\chi_r^2=1.03$ for 55 d.o.f.). For the latter model, which gives the best fit, the T$_{90}$ fluence was $1.25^{+0.10}_{-0.08}\times10^{-6}$\,\fluence\ (15--150\,keV). The flux of the burst extrapolated to the energy range 0.3--10\,keV was $\approx2\times10^{-9}$\,\flux.

\begin{figure*}
\centering
\includegraphics[width=0.73\textwidth]{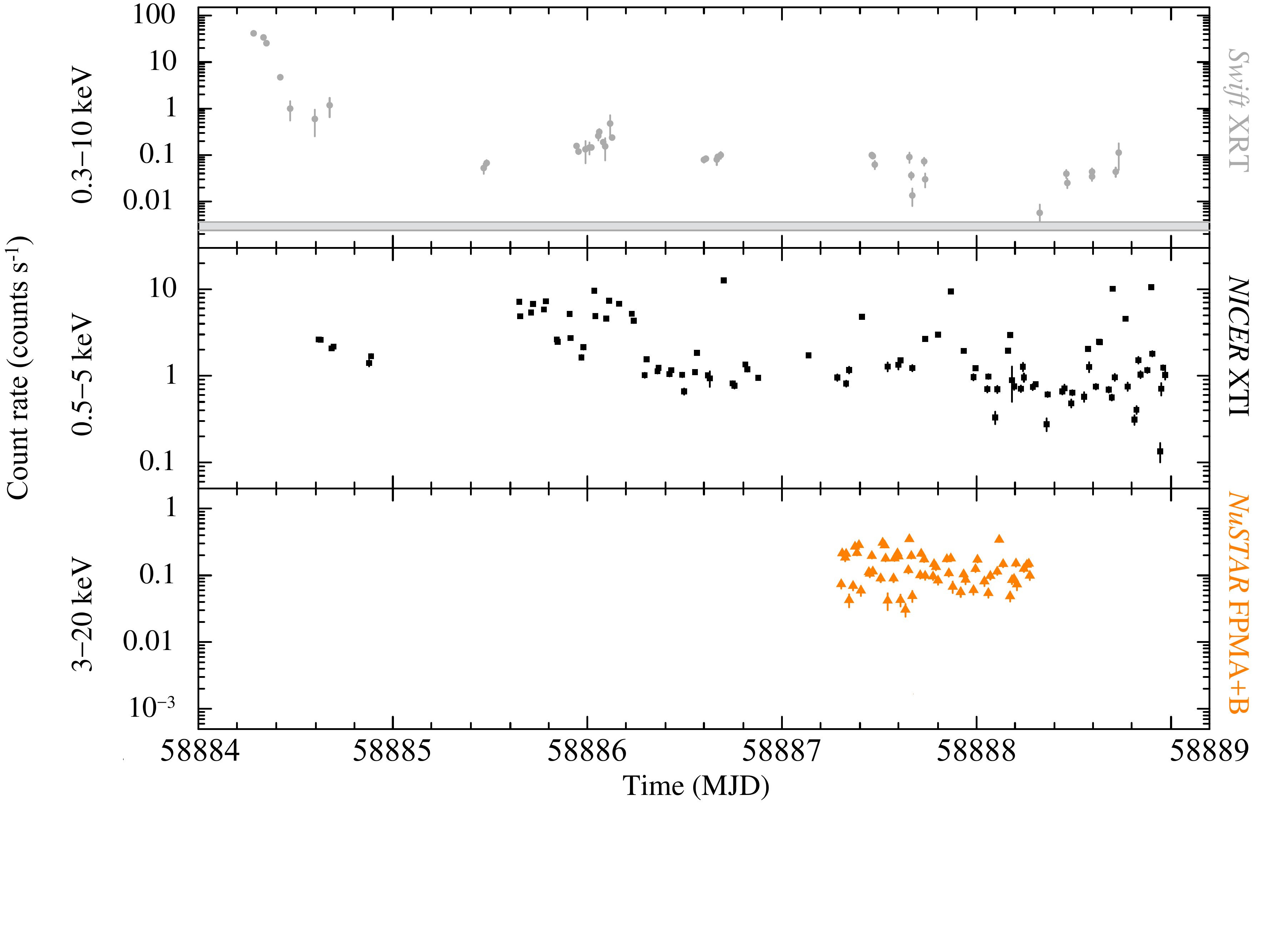}
\vspace{-1.8cm}
\caption{Multi-instrument background-subtracted X-ray light curves of \src\ between 2020 February 5--10, binned at 800\,s. The grey shaded region in the top panel indicates the confidence interval (at 3$\sigma$) for the quiescent \swift/XRT net count rate computed by stacking all observations performed since 2020 February 10 (see Table\,\ref{tab:observations}). In most cases, the marker size is larger than the error bars.}
\label{fig:lc}
\end{figure*}

\begin{figure*}
\centering
\includegraphics[width=0.98\textwidth]{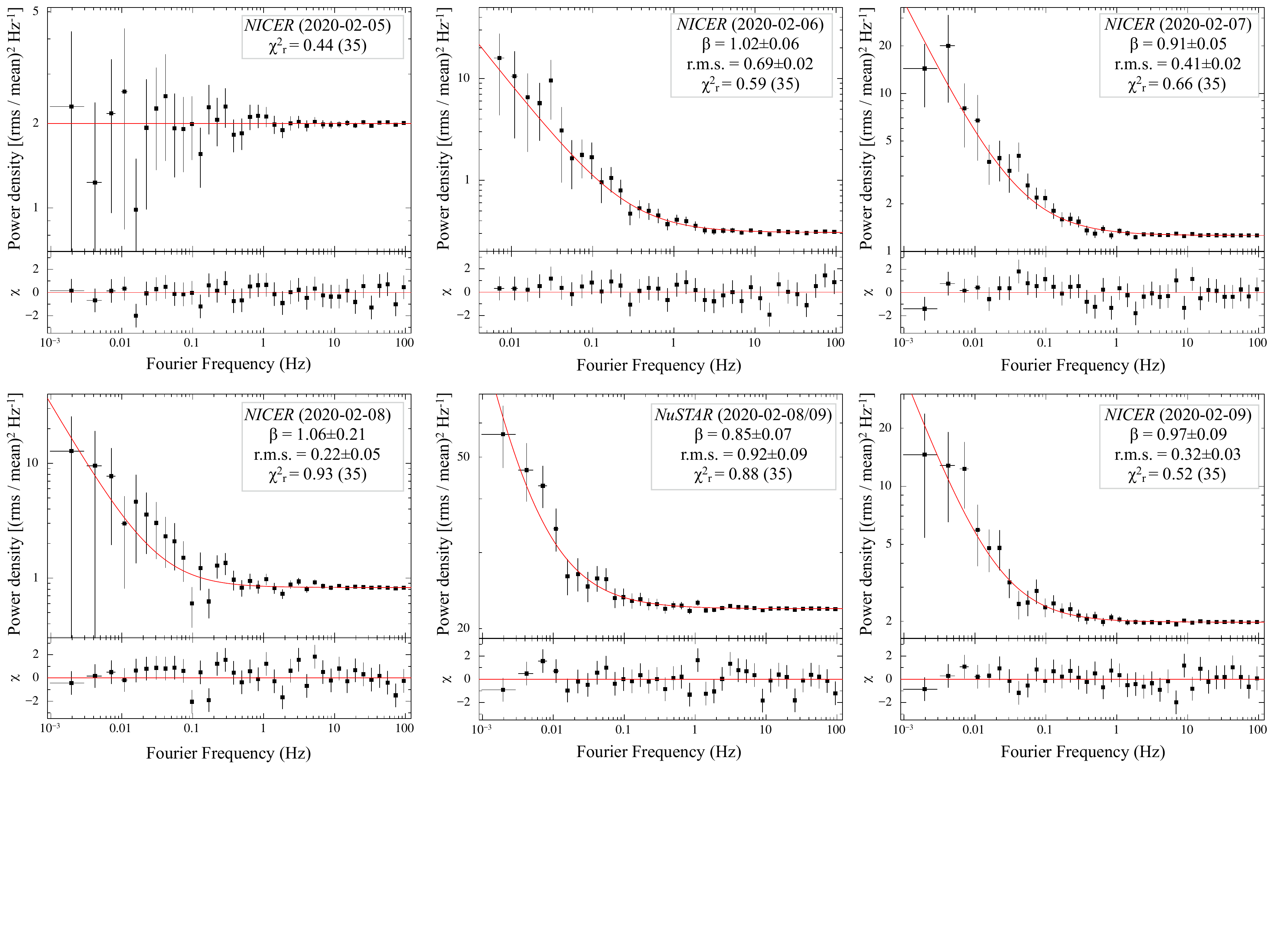}
\vspace{-2.8cm}
\caption{Power density spectra extracted from \nicer\ (0.5--5\,keV) and \nustar\ (3--20\,keV) data. The red solid lines mark the best-fitting model to the power spectra (see text for details). The best-fitting values for the power-law index, the rms variability amplitudes below 1\,Hz and the $\chi_r^2$ values with the d.o.f. of the fits are also shown. Post-fit residuals are shown in the bottom panels. Uncertainties are at 1$\sigma$ c.l.}
\label{fig:timing}
\end{figure*}

\subsection{X-ray variability and searches for periodic emission}
\label{sec:timing}

The X-ray time series reveal considerable variability on timescales of minutes in all observations performed in outburst, in the form of drops in the count rates as well as erratic flares (Figure\,\ref{fig:lc}). We checked the \nicer\ and \nustar\ time series for the presence of periodic dips adopting different time bins, but found none. 
To study the aperiodic time variability, we extracted the power density spectra for all \nicer\ and \nustar\ data sets. We sampled the time series of each observation with a time bin of 2$^{-8}$\,s and calculated power spectra into time intervals of length 2$^{9}$\,s using the fractional rms-squared normalization. We then rebinned the resulting average spectrum as a geometric series with a step of 0.3. The power density spectra are shown in Figure\,\ref{fig:timing}. Except for the first \nicer\ observation, the aperiodic variability of \src\ is characterised in all cases by a noise component that increases towards lower frequencies below $\approx$1\,Hz. We modelled the power density spectra using a function of the form $P(\nu) = K + C\nu^{-\beta}$, accounting for the white noise (first term) and red noise (second term) components. The best-fitting values for the power-law index and the fractional r.m.s. variability amplitudes evaluated at frequencies below 1\,Hz for all observations are shown in Figure\,\ref{fig:timing}. We observe large values of the variability amplitude, up to $0.92\pm0.09$ in the \nustar\ observation. Similar values were derived when averaging power spectra computed over shorter time lengths (down to 2$^7$\,s), and/or when applying different geometrical rebinning factors to the averaged spectrum.
We found no evidence for variability features over restricted frequency ranges (e.g. quasi-periodic oscillations) in any of the power spectra.

\begin{figure}
\centering
\includegraphics[width=0.45\textwidth]{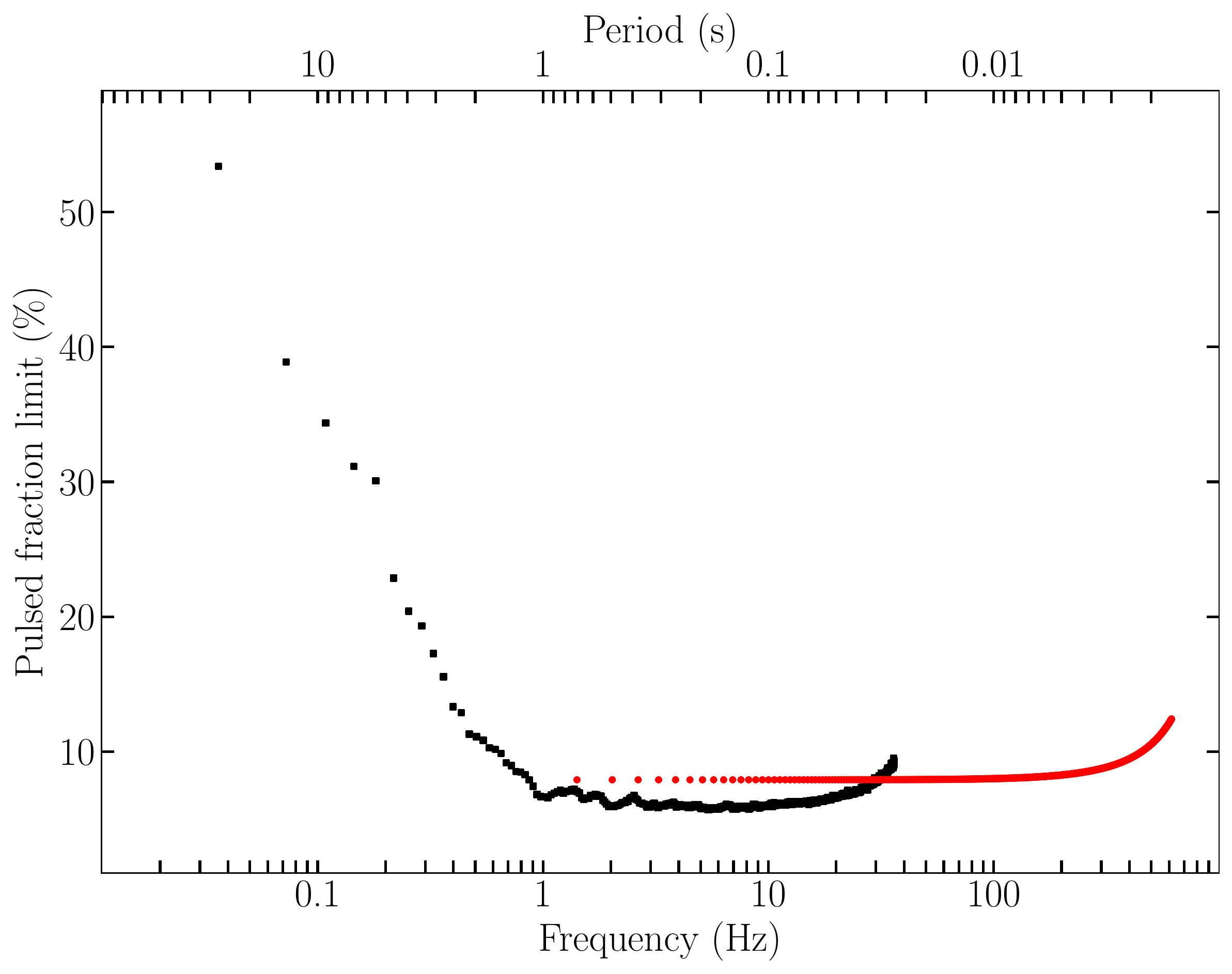}
\caption{3$\sigma$ upper limits on the pulsed fraction for any coherent signal within the frequency range 0.01--620\,Hz, estimated by means of accelerated searches on the combined \nicer\ datasets over the 0.5--5\,keV energy range (see Section\,\ref{sec:timing} for details).}
\label{fig:pful}
\end{figure}

We searched for coherent signals in the \nicer\ light curves in the 0.5--5\,keV energy range by adopting the recipe outlined by \citet{israel96} and taking into account also the effects of signal smearing introduced by the presence of a first period derivative component $\dot{P}$ induced by an orbital motion. We corrected the photon arrival times by a factor $-\frac{1}{2}\frac{\dot{P}}{P}\,t^2$ for a grid of about 1800 points in the range 5$\times$10$^{-6}<|\frac{\dot{P}}{P}\,($s$^{-1})|<$10$^{-11}$  (see \citealt{guil20}  for details). No significant peak was found. Figure\,\ref{fig:pful} shows the 3$\sigma$ upper limits to the pulsed fraction (defined as the semi-amplitude of the sinusoid divided by the average count rate) obtained in two cases: maximizing the Fourier resolution $T_{\rm obs}^{-1}$ (where $T_{\rm obs}$ is the observation length; black points), and extending the search down to periods of the order of milliseconds (red points). In the best cases, we obtained upper limits around 8--12\% and 6--7\% in the period ranges 1.6--50\,ms and 50\,ms--1\,s, respectively. 

We also inspected the \swift/XRT data acquired in quiescence for the presence of a possible modulation of the soft X-ray emission at periods in the range from days to months. We found no evidence for any periodicity on these timescales.

\subsection{X-ray spectrum in outburst and spectral evolution}
\label{sec:spectra}
We modelled the broad-band spectrum extracted using quasi-simultaneous data acquired by \nustar\ and \swift/XRT (obs.ID: 00954304004). We included 
a renormalization factor in the modelling to account for intercalibration uncertainties, as well as for possible differences in the X-ray flux registered by the two observatories (\nustar\ continued observing the source for $\sim$13.5\,hr after the end of the \swift/XRT observation, over a time interval of significant X-ray variability; see  Table\,\ref{tab:observations} and Figure\,\ref{fig:lc}). For all the models tested, the correction was always smaller than 10\%. Single-component models such as a blackbody, a bremsstrahlung, a multicolour blackbody emission model from an accretion disk ({\sc diskbb} in {\sc xspec}), or optically thin thermal plasma emission models ({\sc mekal} and {\sc apec}; \citealt{smith2001}) are rejected by the data ($\chi^2_r\geq1.6$ for 99 d.o.f. in all cases). An absorbed power-law model provides instead a more acceptable result ($\chi^2_r=1.16$ for 99 d.o.f.). 
However, based on $F$-tests, we deem that the broad-band spectrum is better described by a double-component model comprising also a thermal component at low energies, such as an absorbed blackbody plus power-law model. We derived $\chi^2_r=1.04$ for 97 d.o.f. and an $F$-test probability of chance improvement of $\sim$2$\times10^{-3}$. The best-fitting parameters are $\nh=4.6^{+1.3}_{-1.0}\times10^{21}$\,cm$^{-2}$, $kT_{\rm BB}=0.35\pm0.06$\,keV, $R_{\rm BB}=2.7_{-0.7}^{+2.1}$\,km (assuming a distance of 10\,kpc), $\Gamma=1.77^{+0.06}_{-0.07}$.
The observed flux over the 0.3--20\,keV energy band was $(3.2\pm0.3) \times10^{-12}$\,\flux, with a fractional contribution of the thermal component of $\simeq$30\% in the same
band.
A double-blackbody model gives instead a much worse description of the data ($\chi^2_r=1.56$ for 97 d.o.f.). 
More sophisticated models are beyond the scope of this work. 

We did not find evidence for the presence of broad emission or absorption features in the spectrum. We included an additional Gaussian component with centroid allowed to vary in the energy range 6.4--6.97\,keV and set 3$\sigma$ upper limits of 40--80\,eV on the equivalent width associated with any iron emission line with widths in the range 0.1--0.4\,keV.

\begin{figure}
\centering
\includegraphics[width=0.45\textwidth]{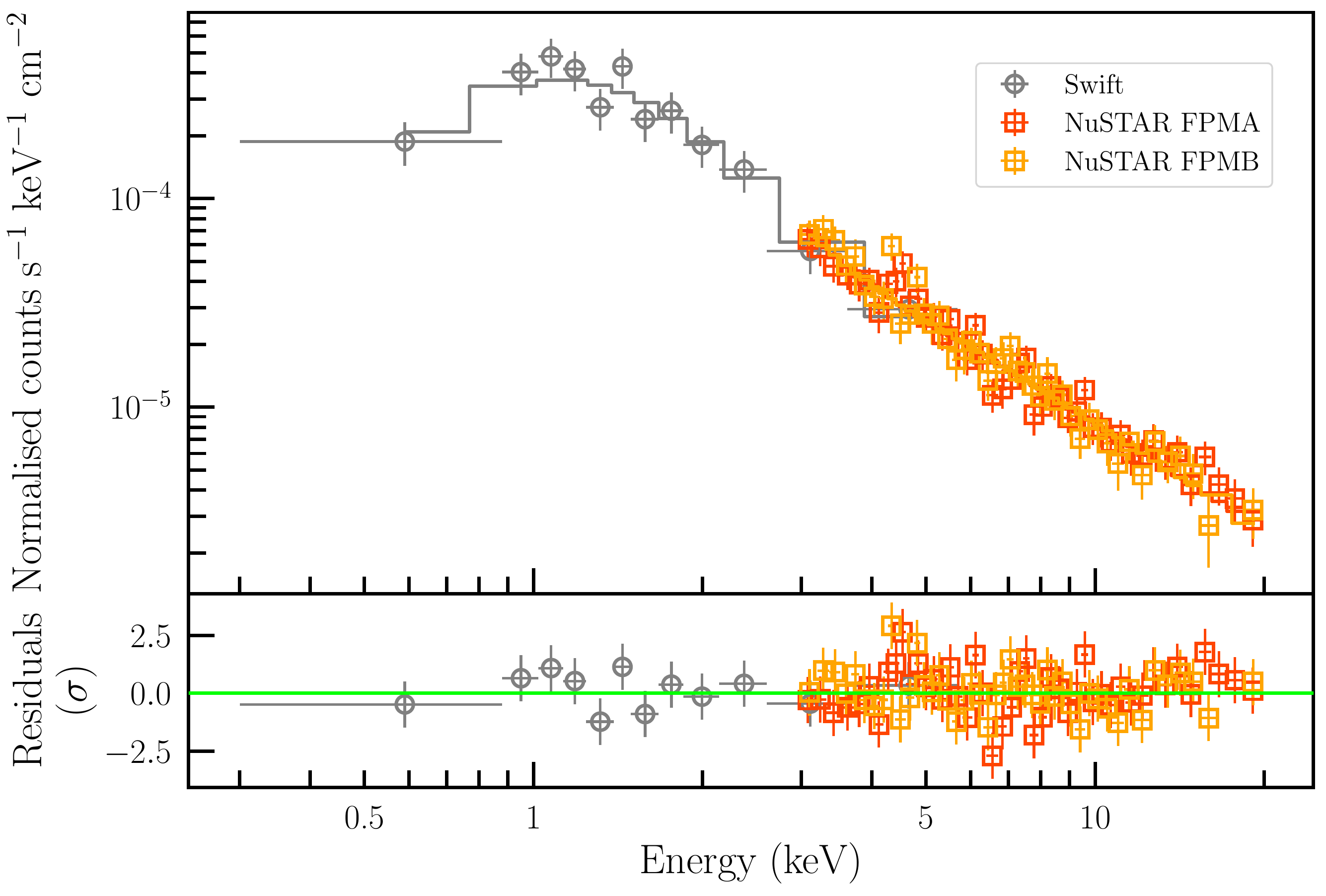}
\caption{Broad-band X-ray spectrum of \src\ in outburst, extracted from quasi-simultaneous \swift/XRT and \nustar\ data. The best-fitting blackbody plus power-law model corrected for absorption is indicated using the solid line. Post-fit residuals are also shown in the bottom panel.}
\label{fig:spectra}
\end{figure}

To assess the overall evolution of the source spectrum and flux along the outburst decay, we then fit the absorbed BB$+$PL model to the \swift/XRT spectra extracted separately from each observation. The limited energy range covered by the XRT together with the relative scarce photon counting statistics at high energies do not allow us to track in detail the evolution of the thermal and nonthermal spectral components. In the following, we assume that the power-law component is present all along the outburst decay down to quiescence and that its slope does not vary in time (see Section\,\ref{sec:discussion} for a more in-depth discussion on this assumption). For the joint fits, we thus fixed the photon index to the value derived from the analysis of the broad-band spectrum, $\Gamma=1.77$. The column density was also held fixed to the above value, $\nh=4.6\times10^{21}$\,cm$^{-2}$. All other parameters were allowed to vary. We obtained $\chi^2_r=1.08$ for 674 d.o.f. 

The results of the spectral analysis are reported in Table\,\ref{tab:observations}. 
We observe a clear overall softening in the spectrum, with the blackbody temperature decreasing from an initial value of $\sim$1.5\,keV at the outburst peak down to $\sim$0.1\,keV in quiescence.
The observed flux rapidly decayed from $\sim1.8\times10^{-9}$ to $\sim3\times10^{-11}$\,\flux\ over the first 3\,hr after the trigger, and then down to a steady value of $\sim1.3\times10^{-13}$\,\flux\ over the subsequent few days (here and in the following, all fluxes are quoted in the 0.3--10\,keV energy range). \src\ has been lingering at such X-ray flux level since then. 

The field of the source was also covered by \xmm\ slew observations eight times between November 2004 and May 2019, before the outburst onset. \src\ was undetected in all these pointings. The deepest limit on the count rate at the source position, $<$0.9 counts\,s$^{-1}$ (3$\sigma$; 0.2--12\,keV), translates into an observed flux of $<2\times10^{-12}$\,\flux, assuming the \swift/XRT quiescent spectrum. This limit is compatible with the quiescent flux measured in the stacked \swift/XRT observations performed since 2020 February 10.

 \subsection{The optical and UV emission}
 \label{sec:optuv}
 
 \subsubsection{Photometry}
 \label{sec:photom}
 
The UV counterpart was detected using the \swift/UVOT at a magnitude of $\sim$17 soon after the burst trigger (here and in the following, magnitudes are reported in the Vega system). It decayed to $\sim$18.5 a few hours later, then down to $\sim$20.3 on February 6 and faded below the instrument detection sensitivity in exposures taken over the next days and from then on (with typical 5$\sigma$ upper limits of $>$20.5 in 4-ks exposures).
To assess whether the emission decay pattern in the UV band is similar to that observed in the X-rays, we first attempted fitting the time evolution of the XRT count rates (averaged over chunks of variable lengths) and of the UVOT fluxes (in the single images) using different simple models, such as one or two power laws, a power law plus a linear term, a broken power law, one or two exponential functions and a smoothed fast-rise exponential-decay model. However, none of these models is able to capture the large stochastic flux variability of \src, yielding statistically unacceptable fits. To obtain a rough estimate of the decay timescales, we can evaluate the overall flux variation observed over the first two days of the outburst, $\Delta \equiv \Delta F/ \Delta t$, where $\Delta F \equiv F_{\rm max} / F_{\rm min}$. We obtain $\Delta_{\rm X} \approx 3.5$\,hr$^{-1}$ and $\Delta_{\rm UV} \approx 0.45$\,hr$^{-1}$. This rough estimate implies that, during the earliest outburst phases, the flux decayed at a rate that is a factor of $\approx$8 faster in the X-rays than in the UV band.

\begin{figure*}
\centering
\includegraphics[width=0.98\textwidth]{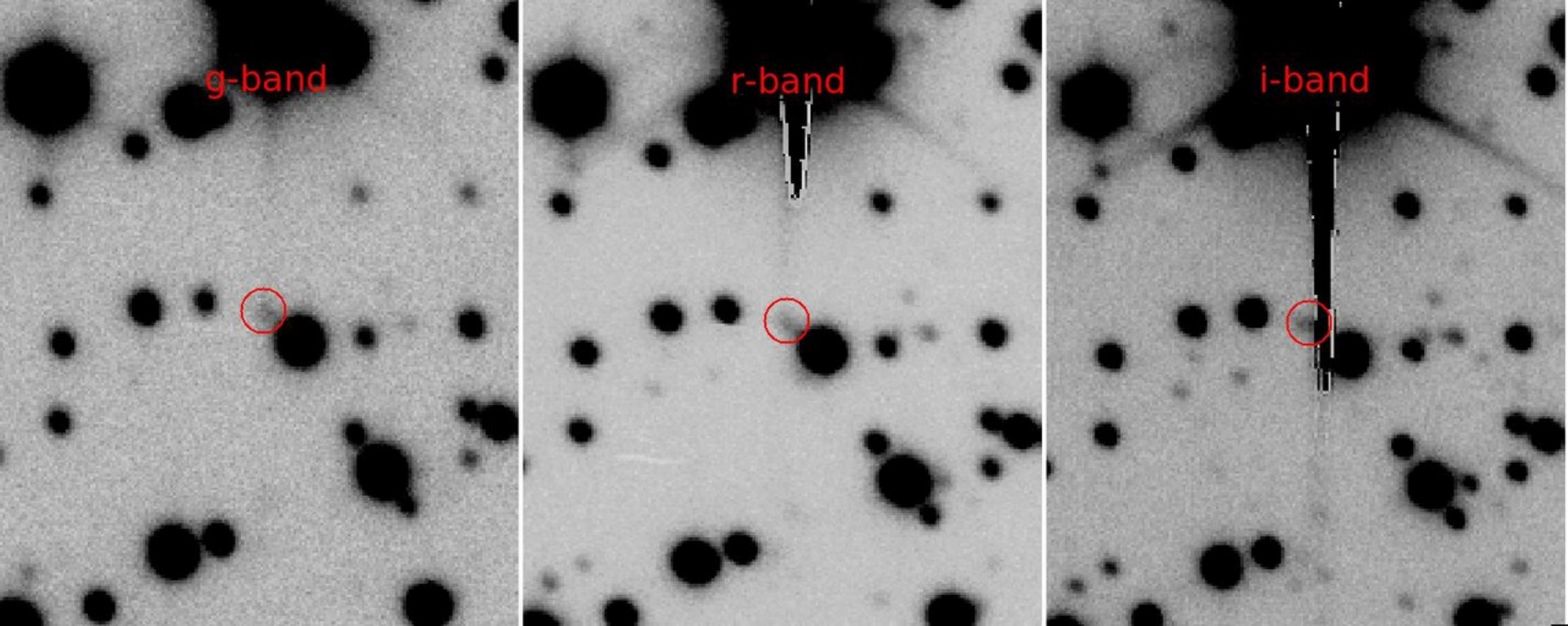}
\includegraphics[width=0.98\textwidth]{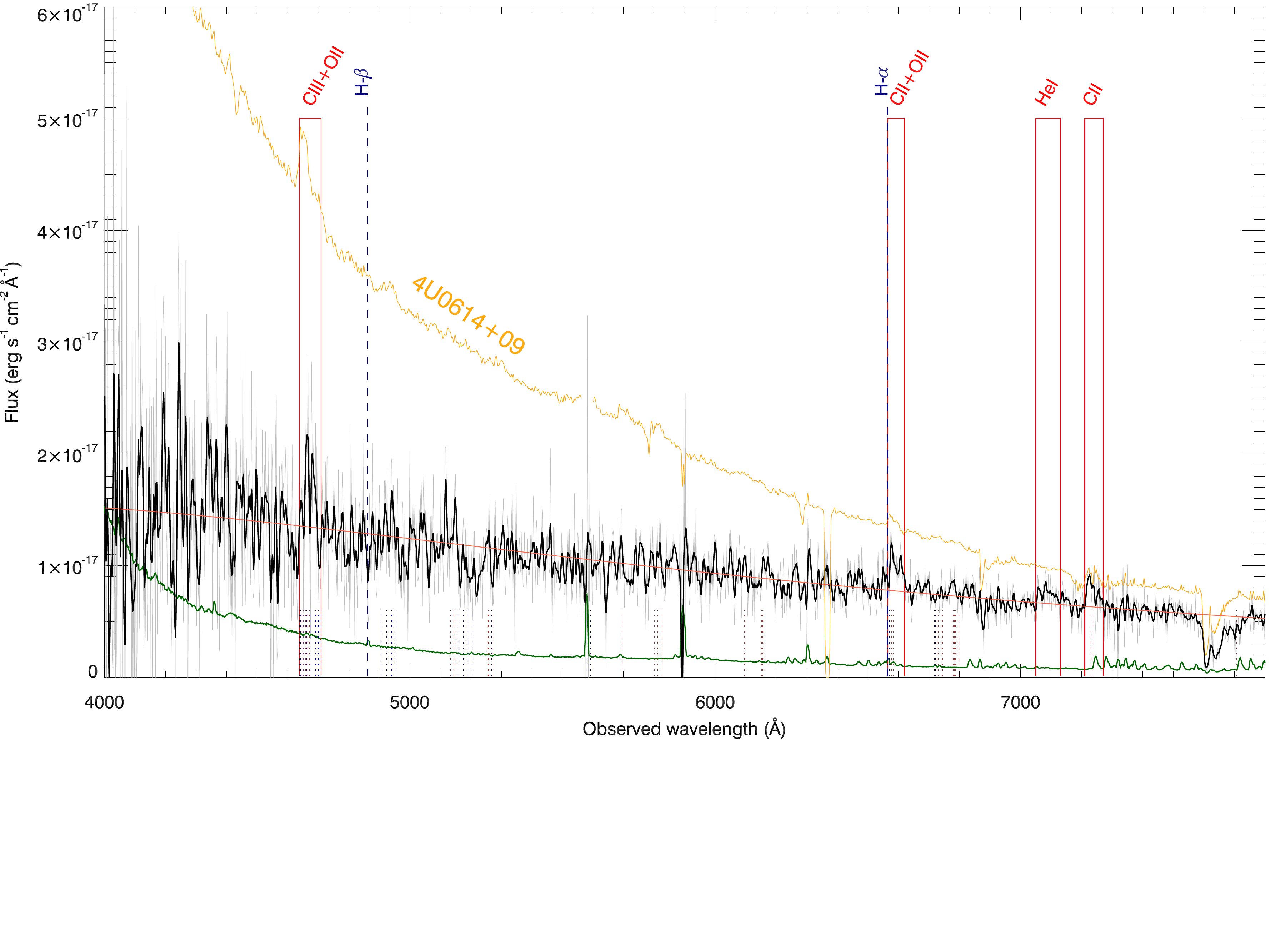}
\vspace{-2.8cm}
\caption{{\em Top}: Optical images acquired by GTC/OSIRIS on 2020 February 18. Saturation blooming due to a nearby star is evident 
in the $i$-band image. The position of \src\ is marked with a red circle. North is up, east to the left. {\em Bottom}: Extinction-corrected optical spectrum of \src\ in quiescence plotted together with a spectrum of the UCXB 4U\,0614$+$091 acquired in archival observations with VLT/X-Shooter. The original data are indicated in grey, the same data convolved with a Gaussian function of width 5\,\AA\ are marked in black. The best fitting blackbody model for the continuum of \src\ is overplotted with an orange line. The most prominent emission features of \src\ are labelled in red. The wavelengths of the H$\alpha$ and H$\beta$ lines are marked in blue. The carbon and oxygen lines identified by \citet{nelemans2004} are marked using brown and blue dotted lines. The broad absorption feature at $\sim$7600 \AA\ is due to the telluric (atmospheric) O$_{\rm 2}$ absorption feature.}
\label{fig:gtc}
\end{figure*}

The optical counterpart was detected at magnitudes $r\simeq16.3$ about 20\,min after the BAT burst \citep{melandri20} and $r\simeq17.5$ about 2.3\,hr later \citep{malesani20}.
The top panel of Figure\,\ref{fig:gtc} shows the calibrated optical images acquired using GTC/OSIRIS about two weeks later, when the source had already returned to quiescence according to the X-ray observations (see Table\,\ref{tab:photometry}). The source appears slightly blended with a nearby star, which may affect the reliability of the photometry. The $i$-band image is also affected by a blooming spike from a bright red source located about 25\,arcsec north. \src\ is detected at an average magnitude $r\simeq21$ in these images.  

From the relation between absorption column density and optical extinction by \citet{foight16} and the $\nh = 4.6 \times 10^{21}$\,cm$^{-2}$ 
derived from the spectral fits of the persistent emission, we can estimate $A_V\sim1.60$, $E(B-V)\sim0.52$. From the extinction curves by \citet{fitzpatrick19}, we derive $A_{UVW2}\sim4.8$ (at 2086\,\AA), $A_{r'}\sim1.3$ (at 6332\,\AA). 
At the outburst peak, the de-reddened UV and optical magnitudes are then $UVW2\sim$12.2 and $r\sim$15.0, respectively, 
giving de-absorbed X-ray-to-UV/optical flux ratios of $F_{\mathrm{X}} / F_{\rm UV} \approx15$ and $F_{\mathrm{X}} / F_{\rm OPT} \approx120$. 
In quiescence, the de-reddened optical magnitude of $r\approx$20 yields comparable X-ray and optical fluxes, $F_{\mathrm{X}} / F_{\rm OPT} \approx2$.

\subsubsection{Spectroscopy}

The bottom panel of Figure\,\ref{fig:gtc} shows the flux-calibrated optical spectrum of \src\ in quiescence, corrected for extinction ($E[B-V]\sim0.52$; see Section\,\ref{sec:photom}). The spectral shape of the continuum emission of \src\ is well fit by a blackbody model with a temperature of $\approx$\,8000\,K. On top of this continuum, we detect a few broad features in the following ranges: 4640--4710\,\AA, which we identify as a blend of multiple lines due to \ion{C}{iii} and \ion{O}{ii}; 6565--6620\,\AA, which we attribute to a blend of \ion{C}{ii} and \ion{O}{ii} lines; and 7210--7270\,\AA\ (\ion{C}{ii}) (see \citealt{nelemans2004}). The profiles of these features are very similar to those detected in the UCXB 4U\,0614$+$091 (Figure\,\ref{fig:gtc}). A weaker broad feature is also seen in the range 7050--7130\,\AA, which we tentatively associate with a blend of lines around \ion{He}{i} at 7065\,\AA. The equivalent widths of these features are reported in Table\,\ref{tab:lines}. We set 3$\sigma$ upper limits on the equivalent width of the H$\alpha$ line in the range from -1.3 to -6.2\AA, assuming line widths in the range 100--500\,km\,s$^{-1}$.

\begin{table}
\caption{
\label{tab:lines}
Identification and equivalent widths of the most prominent optical emission lines of \src.}
\centering
\begin{tabular}{ccc}
\hline\hline
Line	&Wavelength range	&EW	\\
	& (\AA)			&(\AA) \\
\hline
\ion{C}{iii}$+$\ion{O}{ii} & 4638--4710 & -18$\pm$3 \\
\ion{C}{ii}$+$\ion{O}{ii}  & 6565--6620 & -15.6$\pm$1.7 \\
\ion{He}{i}	       & 7050--7130 & -13.0$\pm$1.7 \\
\ion{C}{ii}	       & 7210--7270 & -15.2$\pm$1.5 \\
\hline
\end{tabular}
\end{table}

 \subsection{The radio non-detection in quiescence}

\begin{figure}
\includegraphics[width=0.6\textwidth]{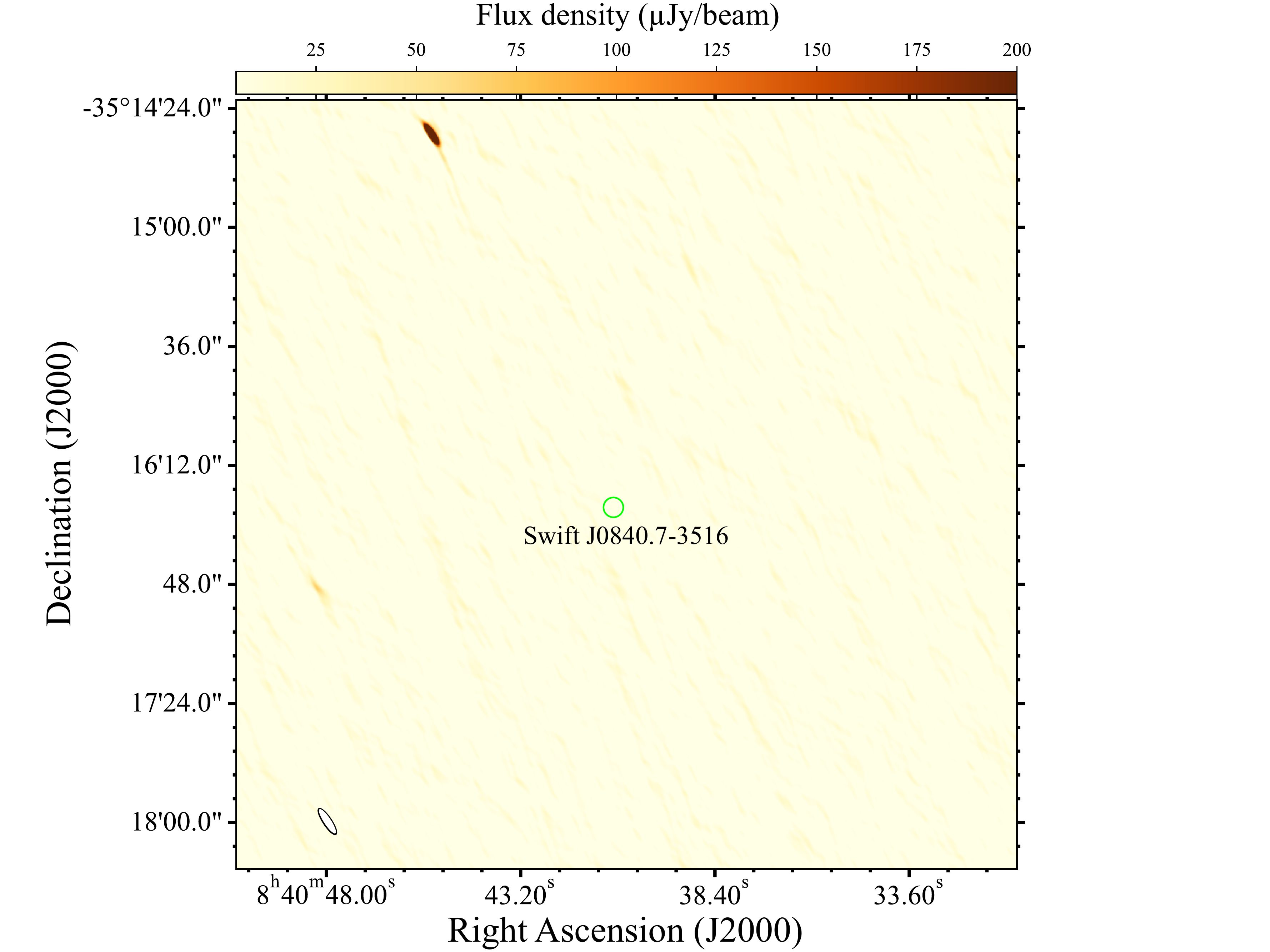}
\caption{Radio image of the field around \src. 
The position of \src\ is marked using a green
circle with an enlarged radius of 3 arcsec for displaying purpose.
The synthesised beam is shown in the bottom left corner of the panel.}
\label{fig:atca}
\end{figure}

During the low-luminosity quiescence phase, we did not detect a radio counterpart at the position of \src\ in our ATCA observations, taken on 2020 February 11 (see Figure\,\ref{fig:atca} for the radio image of the field around \src). We measured 3-$\sigma$ upper-limits on the radio flux density of 27\,$\mu$Jy/beam and 30\,$\mu$Jy/beam at 5.5 and 9\,GHz, respectively. Combining the two observing bands together provided a 3-$\sigma$ upper limit of 18\,$\mu$Jy/beam at a central frequency of 7.25\,GHz.\\

\section{Discussion}
\label{sec:discussion}

\subsection{An ultracompact X-ray binary candidate}

In the following, we discuss possible scenarios for the nature of \src, and show that the observed phenomenology is consistent with a classification as a transient UCXB.  

The rapid luminosity decay observed in \src\ argues against a classification as a tidal disruption event (TDE). The luminosity of TDEs, in fact, declines typically over a much more extended timescale of months to years, in many cases at a rate that is shallower than the canonical power law of the form $L(t) \propto t^{-5/3}$ predicted by standard theories for TDEs \citep{auchettl17}. The above properties point instead to a binary system in our Galaxy. \src\ could be then a magnetic cataclysmic variable (mCV), a high-mass X-ray binary (HMXB), or an LMXB.

The mCVs are systems where a magnetic white dwarf ($B_{{\rm WD}}\simeq10^6-10^7$\,G) accretes matter from a late-type, low-mass star that overflows its Roche lobe (for reviews, see \citealt{mukai2017,demartino2020}). The X-ray spectrum of mCVs consists typically of multi-temperature, optically-thin thermal plasma emission produced in the accretion columns, accounting for the iron complex that is often observed in these systems in the energy range 6--7\,keV. An ubiquitous feature in the spectra of mCVs is the broad iron fluorescent line at 6.4\,keV, with typical equivalent widths in the range $\sim$100--250\,eV. In many cases, an additional soft thermal component from the heated polar cap of the white dwarf has been detected, with a typical blackbody temperature $<$100\,eV. In this respect, the broad-band spectrum of \src\ in outburst is clearly different from what expected from mCVs. We thus consider unlikely \src\ to be a mCV.

The HMXBs are systems where a compact object accretes matter from a massive ($\gtrsim10$\,$M_\odot$) early-type (O or B) donor star (for a review, see e.g. \citealt{reig2011}). The continuum emission of \src\ in the optical band can be adequately described by a blackbody model with a temperature of about 8000\,K. If such emission is indeed dominated by radiation from the donor star (as observed in HMXBs), then the above value for the temperature would be more consistent with an A-type star rather than with an O or B-type star.
The large increase observed in the optical magnitude of \src\ from quiescence to outburst (see above) would be also at odds with what is seen in transient HMXBs, where the bulk of the optical/UV emission is provided in the form of steady radiation by the donor both in outburst and in quiescence, with only a modest variable contribution from an accretion disk.

We then discuss the scenario of a LMXB. The shape of the optical spectrum of \src\ together with the presence of emission features in quiescence argues against a classification as a symbiotic X-ray binary (SyXB), where a NS accretes mass from the stellar wind of an evolved M-type giant donor star (e.g. \citealt{masetti2006,masetti2007}).\footnote{Strong emission features have been detected in one SyXB, \object{GX\,1$+$4}, but only at high X-ray luminosities ($L_x \simeq 10^{36}-10^{37}$\,\lum), that is, in a range where spectral features produced by accretion are expected to emerge over the contribution of the donor (see, e.g. \citealt{chakrabarty97}). The optical emission features in \src\ were detected instead at a much lower X-ray luminosity of $L_X\approx5\times10^{33}d_{10}$\,\lum, where $d_{10}$ is the distance to the source in units of 10\,kpc.} 
This then leaves the scenario in which \src\ is a LMXB where the compact object accretes from the donor via Roche-lobe overflow.

A key diagnostic on the nature of \src\ is provided by its optical spectrum in quiescence. This spectrum shows no sign for the presence of hydrogen emission lines, as instead commonly observed in LMXBs with a hydrogen-rich main sequence donor star. On the other hand, it displays broad emission lines that are generally associated with partially ionised carbon (\ion{C}{ii} and \ion{C}{iii}) and oxygen (\ion{O}{ii}). This phenomenology suggests that an accretion disk almost purely made out of carbon and oxygen was still present around the accretor soon after the end of the outburst.  In this respect, the optical spectrum of \src\ closely resembles that of the persistent UCXBs \object{4U\,0614$+$091}, \object{4U\,1543$-$624} and \object{4U\,1626$-$67} (\citealt{nelemans2004,nelemans2006}; see also Figure\,\ref{fig:gtc}). Based on the strong analogy with the above systems, we propose that \src\ is a new transient UCXB candidate with a carbon-oxygen white dwarf as donor star.

The short duration of the outburst from \src\ makes this system quite peculiar among transient LMXBs. In fact, similarly short outbursts have been detected only in a handful of other cases (e.g. \citealt{heinke10,degenaar14,matasanchez17}).

\subsubsection{X-ray emission mechanisms}

The broad-band X-ray spectral shape of \src\ (see Section\,\ref{sec:spectra}) is typical of LMXBs in outburst, and may be interpreted in terms of repeated inverse Compton up-scattering of soft thermal photons onto a population of hot thermal electrons. As discussed in more detail in Section\,\ref{sec:accretor}, we are unable to identify conclusively the nature of the accretor (BH or NS) with current data. However, in either case, the main production sites of the thermal photons are likely to be the inner regions of the optically-thick, geometrically-thin accretion disk (see \citealt{shakura1973}).  In the case of a NS accretor, additional contributions to the thermal X-ray emission may be provided by the accretion-heated star surface (see, e.g. \citealt{zampieri1995,wijnands2015}) and/or the so-called `boundary layer' that forms when the disk extends all the way down to the NS such that the in-flowing material spreads out over the star surface. However, due to the available photon counting statistics, we are not able to discern the possible contribution of multiple components to the observed thermal emission.
The location and geometry of the hot thermal electrons that Compton up-scatter the thermal photons is instead debated at present: they may be distributed in an extended cloud above the disk (the so-called `corona'; see e.g. \citealt{kara19} and references therein) or in a hot inner flow close to the accretor (e.g. \citealt{done07}).
 
The overall spectral softening observed in \src\ along its outburst decay may be explained in broad terms by the disk instability model for transient UCXBs (and LMXBs in general; see e.g. \citealt{hameury2016}): the decrease in the temperature reflects the gradual transition of the disk from a hot ionised state to a colder neutral state as the mass transfer rate from the donor star decreases and the system approaches quiescence. The large increase of the X-ray flux observed in \src\ might suggest that a large portion of the disk is brought in the hot state in this system \citep{hameury2016}. This process, in fact, should be particularly efficient in UCXBs, since only a relatively small-sized disk can physically fit into their tight binary orbits. In the case of a NS accretor for \src, an additional contribution to the observed spectral softening might be ascribed to low-level accretion onto the star surface. We note, however, that the scarce photon counting statistics currently available in quiescence from the \swift/XRT data ($\sim$540 net counts) precludes a detailed investigation of the evolution of both spectral components throughout the outburst and down to quiescence. In this respect, our assumption of an unchanged power-law slope is certainly simplistic. 

\subsubsection{On the lack of the iron line in the X-ray spectrum}

The absence of the iron K$\alpha$ fluorescent line in the X-ray spectrum of \src\ taken a few days after the outburst onset can be tentatively used
as a tracer of the chemical composition of the disk and donor star, and may provide further support to our classification as a UCXB. In many LMXBs, X-ray photons intercepting the inner regions of the disk are absorbed by highly ionised elements in the disk and re-emitted along the line of sight, producing an additional emission component known as X-ray reflection. The reprocessed emission shows multiple emission features superimposed to the continuum at energies corresponding to transitions in the highly ionised atomic species. Their profiles are broadened and skewed by rotation of the disk material as well as strong Doppler, special relativistic and general relativistic effects inside the gravitational well of the compact accretor \citep{fabian1989}. On the one hand, a broad fluorescent Fe K$\alpha$ line is often detected in LMXBs with hydrogen-rich donor stars (see e.g. \citealt{cackett2010}). On the other hand, this feature is expected to be significantly attenuated in UCXBs with an anomalous abundance of carbon and oxygen, owing to the screening of the presence of iron and other heavy elements in the carbon-oxygen-dominated disk material (\citealt{koliopanos2021} and references therein; see also \citealt{zand2007}). Indeed, \citet{koliopanos2021} found no evidence of emission associated with the iron line in the X-ray spectra of seven known UCXBs, with upper limits on the equivalent widths of 8--60\,eV (3$\sigma$). Based on these values, they estimated an oxygen-to-iron ratio which is at least an order of magnitude larger than the Solar value, and ascribed this property to the presence of a carbon-oxygen (or even oxygen-neon-magnesium) donor star. In this framework, the upper limits inferred on the equivalent width of any iron line in \src, 40--80\,eV (see Section\,\ref{sec:spectra}), although not as constraining as the limits derived by \citet{koliopanos2021} for other UCXBs, are anyway smaller than the values observed in many hydrogen-rich LMXBs, and still support a classification of \src\ as a UCXB with a carbon-oxygen white dwarf. We stress, however, that important exceptions exist to the arguments outlined above, since evidence for iron emission lines has been actually found in a few UCXBs with possibly carbon-oxygen white dwarf donors (e.g. \citealt{vandeneijnden18,ludlam19,ludlam20}). Alternatively, the lack of the iron emission line in \src\ may be related to an increase in the ionization of the inner accretion flow towards lower mass accretion rates, similarly to what has been proposed by \citet{vandeneijnden20} for the case of 4U\,1608$-$52.

\subsubsection{Optical and UV properties and an orbital period estimate}
\label{sec:optporb}

The enhanced UV and optical emission detected at the outburst peak decayed at a rate that is a factor of $\approx$8 slower than that observed in the X-ray band during the earliest outburst phases. This emission component may be interpreted in terms of enhanced irradiation of the outer regions of the disk and the donor (in analogy with most LMXBs). The optical counterpart is much fainter and blue as soon as the system entered in quiescence, suggesting that the optical emission is still dominated by the disk in this phase.
Assuming a distance in the range 5--10\,kpc (see Section\,\ref{sec:accretor}), the dereddened magnitude of \src\ in quiescence translates into an absolute magnitude for the donor star of $M_{r'}>6.1$ (for $D=5$\,kpc) or $M_{r'}>4.6$ (for $D=10$\,kpc), where the values should be considered as upper limits owing to the unknown contribution of the disk and the donor star to the optical emission in quiescence. These values make the donor star of \src\ among the faintest in the population of known LMXBs, and are similar to those estimated for other UCXBs (e.g. \citealt{bassa06}).

Using the empirical relation between the absolute magnitude of a LMXB, the binary orbital period and the X-ray luminosity by \citet{vanparadijs1994}, adopting the value for the Eddington luminosity for a NS with a mass of $1.4 M_{\odot}$ ($L_{\rm Edd} \simeq 2.5 \times 10^{38}$\,\lum) and accounting for the uncertainties on the distance, we can roughly estimate an orbital period of a few tens of minutes for \src. This is well in the range of those measured for UCXBs.

\subsection{The nature of the accretor}
\label{sec:accretor}

Determining the nature of the compact object in an unclassified LMXB is often challenging. It can be assessed based either on a dynamical measurement of
the mass of the compact object (via optical and near infrared spectroscopy during quiescence; see e.g. \citealt{casares1992}), or on the detection of X-ray coherent pulsations and/or thermonuclear X-ray bursts, which would provide a straightforward observational evidence for a NS accretor. To-date, all confirmed UCXBs are known to contain a NS accretor, although there are a number of BH candidates \citep[see e.g.][]{bahramian2017}. Assuming that our identification of \src\ as a UCXB is correct, it is then tempting to conclude that this system hosts a NS as well. However, our non-detection of coherent pulsations or thermonuclear X-ray bursts leaves this as an open question.

\subsubsection{On the lack of coherent X-ray pulsations}

Coherent pulsations are thought to be formed in NS systems where the magnetic field is large enough to truncate the accretion disk and channel part of the disk material onto a small region on the NS surface, close to the magnetic poles. The radiation emitted from the heated impact region (hot spot) or from a slab of shocked plasma that forms above it (accretion columns) appears modulated at the NS spin period, producing a pulsed emission component. About 30 accreting millisecond X-ray pulsars have been detected to date (for reviews, see \citealt{disalvo2020,patruno2021}) and some of them were seen to exhibit pulsed fractions as small as a few percent in outburst (e.g. \citealt{strohmayer18}), that is, below the upper limits estimated for \src\ (Section\,\ref{sec:timing}). Besides, three accreting NS--LMXBs have shown coherent X-ray pulsations only sporadically during their outbursts, with the most extreme case being represented by Aql X-1, which showed pulsations over a $\sim$150\,s interval out of a total observing time of $\sim$1.5\,Ms \citep{casella2008}. However, most NS--LMXBs are non-pulsating systems, and a systematic analysis by \citet{patruno2018} suggested that weak pulsations might not form at all in (most) non-pulsating LMXBs. Our non-detection of pulsations from \src\ is thus not sufficient to conclusively rule out a NS accretor and might still point either to an accreting millisecond pulsar characterised by a very small X-ray pulsed fraction and/or intermittent pulsations, or a non-pulsating NS--LMXB.

\subsubsection{On the lack of thermonuclear X-ray bursts}

Many NS--LMXBs exhibit thermonuclear (type-I) X-ray bursts due to unstable ignition of hydrogen and/or helium freshly accreted onto the NS surface (for a review, see \citealt{degenaar18}). Bursts triggering all-sky monitors (including \swift/BAT) have been detected from several NS--LMXBs and in some cases they even 
led to the discovery of new transient systems (e.g. \citealt{wijnands09,degenaar12}). These episodes typically last a few tens of seconds, though events as long as a few hours 
have been observed from NSs accreting at low X-ray luminosities (e.g. \citealt{intzand19}). Their spectra are commonly described adequately by a blackbody model with temperatures of $kT_{\rm BB}\simeq2-3$\,keV (see \citealt{galloway2020} and references therein). While the duration of the BAT burst from \src\ is per se not that unusual for thermonuclear bursts, its high blackbody temperature of $kT_{\rm BB}\simeq8-9$\,keV (see Section\,\ref{BAT:analysis}) is clearly inconsistent with a thermonuclear origin. However, our nondetection of thermonuclear bursts from \src\ is not that surprising. Firstly, these events are generally not expected to occur frequently in UCXBs with a carbon-oxygen rich donor star due to the lack of hydrogen and helium in the disk (see e.g. \citealt{koliopanos2021} and references therein), although noteworthy exceptions exist (e.g. \citealt{kuulkers10}). 
Second, their recurrence times are typically long since the build up of sufficient fuel on the star surface to ignite a burst takes time (see e.g. \citealt{degenaar10}), especially at relatively low luminosity levels such as those observed from \src. Therefore, our non-detection of thermonuclear bursts from \src\ is, all in all, not inconsistent with a NS accretor.

\subsubsection{Clues from X-ray spectral properties}
\label{sec:clues}

The X-ray spectral properties can also be used to help establish the nature of the compact object in a LMXB. Indeed, the X-ray spectra of transient NS systems are observed to be significantly softer than those of BH systems at luminosities below 10$^{35}$\,\lum\ (0.5--10\,keV; see \citealt{wijnands2015}). Clearly, this diagnostic tool strongly relies on the distance to the system, which is unknown in the case of \src.\footnote{The optical magnitude of \src\ in quiescence is fainter than the sensitivity limit of the {\em Gaia} mission, and indeed the source is not listed in the {\em Gaia} Early Data Release 3 ({\em Gaia} EDR3; \citealt{gaia20}) and no information on its geometric parallax is available.} Nevertheless, we can derive some constraints on the distance by considering the model for the spiral structure of the Galaxy derived from the distribution map of the \ion{H}{ii} regions within the Galaxy \citep{hou14}. Their distribution and the large absorption column density derived from our spectral analysis ($\nh \simeq 4.6 \times 10^{21}$\,cm$^{-2}$; Section\,\ref{sec:spectra}), which is comparable to the total Galactic column in the direction of \src\ ($\nh \simeq 4.3 \times 10^{21}$\,cm$^{-2}$; \citealt{willingale2013}) suggest that \src\ is located at a distance in the range 5--10\,kpc and likely in the Perseus Arm. 
Fitting again the \swift/XRT spectra taken in outburst with an absorbed PL model and column density allowed to vary shows an increase of the photon index from $\Gamma\simeq1-1.5$ at fluxes $F_{\rm X,unabs}\gtrsim10^{-9}$\,\flux\ (i.e. $L_X\gtrsim10^{37}d^2_{10}$\,\lum) to $\Gamma\simeq2-3$ at fluxes $F_{\rm X,unabs}\lesssim6\times10^{-11}$\,\flux\ (i.e. $L_X\lesssim7\times10^{35}d^2_{10}$\,\lum) (see Table\,\ref{tab:swiftpl}). For these values, the evolution of the \src's position on the photon index versus X-ray luminosity plane presented by \citet{wijnands2015} seems more compatible with the track followed by NS--LMXBs rather than BH--LMXBs (see also \citealt{parikh17}).

Remarkably, the thermal fraction of the total X-ray luminosity \src\ in quiescence ($\sim$46\%) appears to be broadly consistent with that reported for several transient NS--LMXBs accreting at low luminosities. In these systems, the co-existence of a thermal and a non-thermal components has been ascribed to a combination of cooling emission from the
NS that has been heated by mass accreted
in outburst, ongoing low-level accretion onto the NS surface, and/or emission mechanisms associated with the magnetosphere (see e.g. \citealt{jonker04,degenaar2013,campana2014,wijnands2015}). 
We stress, however, that the above value for the thermal fraction of \src\ was estimated assuming no variation in the power-law slope along the outburst. Deep broad-band observations would be needed to characterise adequately the quiescent X-ray spectrum and the spectral energy distribution from the optical to the X-rays. 

\begin{figure}
\includegraphics[width=\columnwidth]{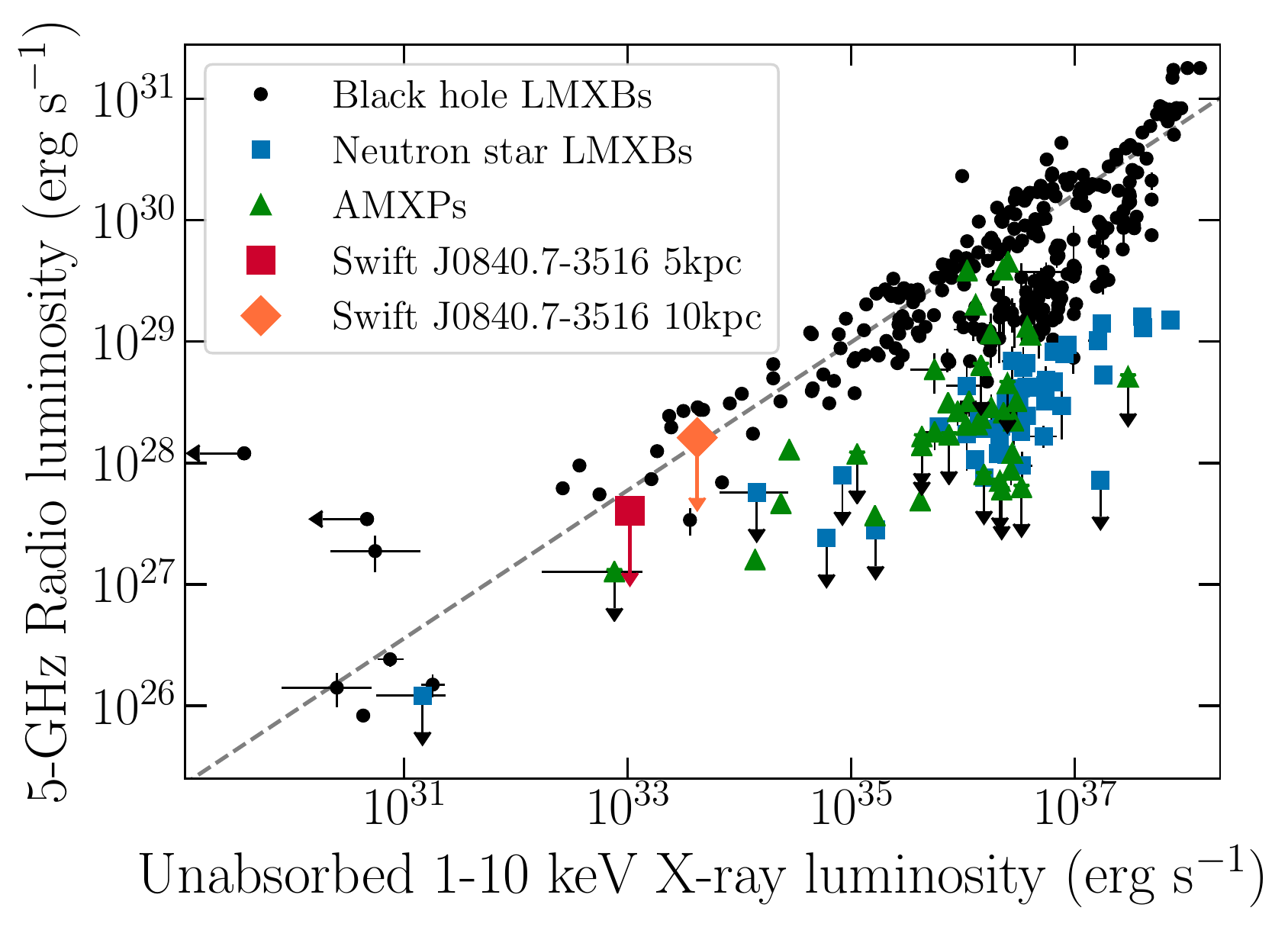}
\caption{Radio and X-ray luminosities of accreting BHs and NSs. The quasi-simultaneous radio and X-ray luminosities of \src\ during quiescence are shown using the red square and orange diamond for the assumed distances of 5 and 10\,kpc, respectively. We show the population of BH--LMXBs (black circles), NS--LMXBs (blue squares), and accreting millisecond X-ray pulsars (AMXPs; green triangles). The grey dashed line indicates the best-fit correlation for the BH systems from \cite{gallo2018}. Our radio and X-ray observations of \src\ do not differentiate between a BH or NS accretor. Data taken from \citet{arash_bahramian_2018_1252036}.}
\label{fig:lrlx}
\end{figure}

\subsubsection{Other constraints}

In their low-luminosity or quiescent hard states, BH and NS--LMXBs exhibit a non-linear relationship between their radio and X-ray luminosities \citep[e.g.][]{hannikainen1998,corbel2000,gallo2003,gallo2012,corbel2013,tudor17,gusinskaia20a}. While the behaviour of single systems may differ somewhat \citep[e.g.][]{tudor17,russell18,vde2020}, analysis of the full population of BH and NS systems has shown that NS--LMXBs are typically a factor of $\sim$22 more radio faint than their BH counterparts \citep{gallo2018}. As such, a source's radio and X-ray brightness has often been used to help discriminate between a NS or BH LMXB. Placing our 3-$\sigma$ radio upper-limit with the closest in-time X-ray luminosity (1-day separation) on the radio -- X-ray luminosity plane (Figure~\ref{fig:lrlx}) shows that our data are consistent with both a BH and a NS LMXB for our assumed distances. Hence, the upper-limit on the radio emission during quiescence does not identify the nature of the accretor. During quiescence, more sensitive radio observations that detect the source, or monitoring during a new outburst may be able to discriminate between the two classes using this method.

As a final remark, we note that for similar binary orbital periods, NS systems are typically brighter in the X-rays than BH systems in quiescence. Taking the quiescent X-ray luminosity of \src\ as a face value ($L_X\sim5\times10^{33}d^2_{10}$\,\lum) and assuming an orbital period of the order of a few tens of minutes (as estimated in Section\,\ref{sec:optporb}), we can place \src\ on the X-ray luminosity vs. orbital period plane and notice that its position would be far more consistent with a NS accretor than with a BH accretor (see e.g. Fig.\,3 by \citealt{armaspadilla14}, and references therein). However, caution should be exerted in overinterpreting this result, since low-level accretion may provide a non-negligible contribution to the emission detected from \src\ (see Section\,\ref{sec:spectra}) and the `genuine' quiescent X-ray luminosity may be actually lower than the one assumed above to some extent.

\section{Conclusions}
\label{sec:conclusions}

\src\ was discovered as it entered a X-ray, UV and optical outburst on 2020 February 5. This transient episode was characterised by an increase in the X-ray flux of a factor of $\approx$10$^4$ above quiescence and lasted overall only $\sim$5 days. The source was detected up to energies of about 20\,keV in outburst, and showed substantial aperiodic time variability on timescales as short as a few seconds all along the active phase. The X-ray spectrum was well described by the superposition of a thermal component and a non-thermal component, and it softened considerably as the source recovered its quiescent state. At the outburst peak, the UV and optical emission reached magnitudes of $\sim$17 and $\sim$16.3, respectively. The UV emission rapidly decayed by about 3 mag in $\sim$1.5\,days and then became no more detectable by \swift, while the optical emission decreased by about 1\,mag in $\sim$2.6\,hr and, overall, by almost 5 mag as the source recovered quiescence. The optical spectrum in quiescence is well described by a blackbody model with a temperature of $\approx$8000\,K, and shows broad emission features mostly associated with ionised carbon and oxygen. 

We discussed several scenarios for the nature of this source, and showed that its phenomenology points to a transient UCXB possibly with a white dwarf donor star. While current data are insufficient to draw a firm conclusion on the nature of the accretor, we favour the scenario of a NS accretor based on the multi-band properties observed both in outburst and in quiescence.

By definition, confirmation of \src\ as an UCXB can only be obtained through the measurement of the orbital period. A method that proved to be successful in relatively low-inclination systems has been the detection of a periodic optical modulation in photometric observations (e.g. \citealt{zhong2011,wang2015}). This arises from X-ray heating of the donor star by the irradiating X-ray source and to the variation of the visible area of the heated face as a function of the orbital phase. Alternatively, an assessment of the orbital period can be achieved 
via time-resolved optical spectroscopy through radial velocity measurements of the broad emission features that trace the orbital motion of the accreting material close to the accretor. Given the optical faintness of \src\ in quiescence, dedicated observations with 8-m class optical telescopes seem warranted to nail down the nature of this system. In turn, the determination of the orbital period would allow to restrict
the parameter space for a more sensitive search for pulsed emission in existing X-ray data, possibly giving tighter constraints on the nature of the accretor.

\begin{acknowledgements}
We thank F.~Harrison and J.~Stevens for approving Target of Opportunity observations with \nustar\ and ATCA in the Director's Discretionary Time. We thank Peter Jonker for comments on the manuscript and the referee for very useful comments that helped improving the presentation and the discussion of the results. FCZ also thanks Andrea Melandri for providing details on the REM observations taken at the outburst peak and Maria Cristina Baglio and Paolo D'Avanzo for useful discussions. This research is based on observations made with the Gran Telescopio Canarias (GTC), installed at the Spanish Observatorio del Roque de los Muchachos of the Instituto de Astrof\'isica de Canarias in the island of La Palma, under Director's Discretionary Time (code GTC2020-142). The data were obtained with OSIRIS, built by a Consortium led by the Instituto de Astrof\'isica de Canarias in collaboration with the Instituto de Astronom\'ia of the Universidad Aut\'onoma de M\'exico. OSIRIS was funded by GRANTECAN and the National Plan of Astronomy and Astrophysics of the Spanish Government. We are indebted to Antonio Cabrera and the GTC staff for their efforts in performing the GTC observations. The NuSTAR mission is a project led by the California Institute of Technology, managed by the Jet Propulsion Laboratory and funded by NASA. The NICER mission operates on the International Space
Station and is funded by NASA. The Australia Telescope Compact Array is part of the Australia Telescope National Facility which is funded by the Australian Government for operation as a National Facility managed by CSIRO. We acknowledge the Gomeroi people as the traditional owners of the ATCA observatory site. 
This research made use of: the NuSTAR Data Analysis Software (NuSTARDAS) jointly developed by the ASI Science Data Center (ASDC) and the California
Institute of Technology; software provided by the High Energy Astrophysics Science Archive Research Center (HEASARC), which is a service of 
the Astrophysics Science Division at NASA/GSFC and the High Energy Astrophysics Division of the Smithsonian Astrophysical Observatory; IRAF, distributed by the National Optical Astronomy Observatory, which is operated by the Association of Universities for Research in Astronomy (AURA) under a cooperative agreement with the National Science Foundation; APLpy \citep{robitaille12}, an open-source plotting package for Python hosted at \url{http://aplpy.github.com}; and Astropy \citep{astropy:2013, astropy:2018}, a community-developed core Python package for Astronomy. We also made use of the following software: CASA \citep{mcmullin07}, HEASOFT v.~6.28, HENDRICS v.~5.1 \citep{bachetti18}, XSPEC v.~12.11.1 \citep{arnaud96}. FCZ and AB are supported by Juan de la Cierva fellowships. FCZ, AB and NR are supported by the ERC Consolidator Grant `MAGNESIA' (nr. 817661) and acknowledge funding from grants SGR2017-1383 and PGC2018-095512-BI00. TDR acknowledges financial contribution from the agreement ASI-INAF n.2017-14-H.0. We thank support from the COST Action `PHAROS' (CA 16124).
\end{acknowledgements}

\bibliographystyle{aa} 
\bibliography{biblio}

\begin{appendix}
\section{Additional fits of the \swift/XRT spectra}
\begin{table*}
\caption{
\label{tab:swiftpl}
Results of the spectral fits of the \swift/XRT data with an absorbed power-law model with all parameters allowed to vary.}
\centering
\begin{tabular}[!H]{ccccccc}
\hline\hline
Instrument\tablefootmark{a}	&Obs.ID							&\nh 			& $\Gamma$	&$F_{X, {\rm obs}}$\tablefootmark{b} 		&$F_{X, {\rm unabs}}$\tablefootmark{b} 	& $\chi_r^2$ (d.o.f.)  	\\
						&								&($\times10^{21}$\,cm$^{-2}$)	&			& \multicolumn{2}{c}{($\times10^{-11}$\,\flux)}   									&  \\
\hline	
\swift/XRT (WT) 			& 00954304000 					& 4.4$\pm$0.8		& 1.03$\pm$0.06	& 214$\pm$5 	& 239$\pm$6 	& 1.28 (114)	\\
\swift/XRT (WT) 			& 00954304001 					& 7.1$\pm$0.3		& 1.50$\pm$0.02	& 139$\pm$2	& 185$\pm$2 	& 1.21 (451)	\\	
\swift/XRT (PC) 			& 00954304001 					& 7$\pm$1			& 2.3$\pm$0.1		& 2.9$\pm$0.1	& 6.2$_{-0.6}^{+0.8}$ 	& 1.08 (36)	\\
\swift/XRT (PC) 			& 00954304002 	 				& 7$\pm$2			& 2.7$\pm$0.4		& 0.38$\pm$0.06	& 1.2$_{-0.3}^{+0.7}$ 	& 0.81 (6)	\\
\swift/XRT (PC) 			& 00954304005 	 				& 4.6$\pm$0.7		& 2.1$\pm$0.1		& 0.83$\pm$0.06 	& 1.4$\pm$0.1 	& 0.55 (23)	\\
\swift/XRT (PC) 			& 00954304003 	 				& 5.4$\pm$0.9		& 2.5$\pm$0.2		& 0.31$\pm$0.03 	& 0.7$\pm$0.1	 & 0.43 (15)	\\
\swift/XRT (PC) 			& 00954304004 	 				& 7$\pm$2			& 2.7$\pm$0.3		& 0.22$\pm$0.03	& 0.7$_{-0.1}^{+0.2}$	& 0.69 (9)	\\
\swift/XRT (PC) 			& 00954304007 	 				& 5$\pm$1		& 2.1$\pm$0.3 	  	& 0.14$\pm$0.12	& 0.25$_{-0.04}^{+0.06}$ & 0.89 (4)	\\
\swift/XRT (PC) 			& 00954304008--067\tablefootmark{c}   	& 2.0$\pm$0.5		& 1.9$\pm$0.1		& 0.012$\pm$0.001 	& 0.016$\pm$0.001	& 1.38 (32)	\\
\hline
\end{tabular}
\newline
\tablefoottext{a}{The instrumental setup is indicated in brackets: PC = photon counting, WT = windowed timing.}
\tablefoottext{b}{All fluxes are in the 0.3--10\,keV energy range.}
\tablefoottext{c}{Datasets of these observations (59 in total) were merged.}
\end{table*}
\end{appendix}

 
\end{document}